\documentclass{webofc}
\usepackage[varg]{txfonts}   

\woctitle{13th International Conference on Meson-Nucleon Physics and
the Structure of the Nucleon}
\begin{document}
\title{Incoherent photoproduction of $\phi$-meson from deuteron \\at low energies }
%
%

\author{Alvin Kiswandhi\inst{1}\fnsep\thanks{\email{alvin@phys.ntu.edu.tw}}\and
YuBing Dong\inst{2}\fnsep\thanks{\email{dongyb@ihep.ac.cn}},
Shin Nan Yang\inst{1}\fnsep\thanks{\email{snyang@phys.ntu.edu.tw}}
}

\institute{Department of Physics, National Taiwan University, Taipei
10617, Taiwan \and
           Institute of High Energy Physics, Chinese Academy of
           Sciences, Beijing, China}

\abstract{ The LEPS and CLAS data of the incoherent photoproduction
of $\phi$ meson from deuteron at low energies are studied with a
model for $\phi$ meson photoproduction from nucleon consisting of
Pomeron, $\pi$, and $\eta$ meson exchanges in the t-channel, and a
postulated resonance, with parameters fitted to recent LEPS data on
$\phi$ production from proton near threshold. The resonance was
introduced to explain an observed bump in the forward differential
cross section. Within impulse approximation, we find that the Fermi
motion, final state interaction, and the resonance excitation all
give important contributions to improve the agreement with data.
However, discrepancies remain. Contributions  from $\phi$ production
via spectator nucleon by other mesons like $\pi, \rho,$ and $\phi$
produced from the first nucleon need to be  calculated in order to
gain insight on the medium effects as well as the existence of the
postulated nucleon resonance. }
\maketitle
%
The $\phi$-meson photoproduction reaction has long been extensively
studied as it involves many interesting physics issues like, Pomeron
$(P)$ exchange,   nondiffractive processes of the pseudoscalar
$(\pi, \eta)$-meson exchanges,   nucleon exchange, nucleon
resonances excitation,  second Pomeron exchange, $t$-channel scalar
meson and glueball exchanges, and $s\bar s$-cluster knockout.
Recently, another intriguing data of this reaction has appeared.
Namely, LEPS collaboration observed a near-threshold bump structure in the
forward differential cross sections of $\phi$ photoproduction on
protons. It has not been possible to
explain it by the processes mentioned above.

We found in \cite{Kiswandhi10} that, with an addition of a resonance
of $(3/2)^-$, with mass $M = 2.10 \pm 0.03$ GeV and width $\Gamma =
0.465 \pm 0.141$ GeV  to the background mechanism which consists of
Pomeron and $(\pi, \eta)$-meson exchanges in $t$-channel, not only
the  peak in the forward differential cross section but also the
$t$-dependence of differential cross section (DCS), $\phi$ meson
decay angular distribution, and the spin density matrix elements can
be well described.

We have set about to extend the above model further to study the
incoherent photoproduction of $\phi$-meson from deuteron at low
energies, where data have recently become available from LEPS
\cite{Chang10,Chang10_DCS} and CLAS \cite{Gao11}. Our purposes are
three-fold. First is to see whether the resonance postulated in our
model for $\phi$ photoproduction would manifest itself in someway in
the case of deuteron target. Next is to learn some information on
the $\phi$ production from neutron. Lastly, we are interested in the
possible medium effects like Fermi motion, final-state interaction
(FSI), and meson rescatterings on the vector meson propagation in
the simplest nuclear target, the deuteron.

In this contribution, we will present our first results within the
impulse approximation with the FSI effects fully taken into account.

Within the impulse approximation (IA) without final state
interaction, the T-matrix of $\gamma d\rightarrow \phi p n$ is a sum
of $T=T^{(p)}+T^{(n)}$, where the superscript $p$ and $n$ denotes
whether the struck nucleon is a proton or neutron. Symbolically,
e.g., one can write
\begin{equation}
 T^{(p)}=T_{\gamma p\rightarrow \phi p} \Psi(\vec{p_n}-\vec{P_d}/2),
 \end{equation}
 where $\Psi$ denotes the deuteron wavefunction; while $\vec{p_n}$ and $\vec{P_d}$ are the three-momentum of the
 neutron and deuteron, respectively. In our calculation,  all the spin dependences of the $\gamma p\rightarrow
 \phi p$ amplitude $T_{\gamma p\rightarrow \phi p}$,  are taken into
 account and the explicit form of $T^{(p)}$ would then read  like,
\begin{eqnarray}
\hspace{-0.4cm}  {T} ^{(p)}\left(m_\gamma, m_d; m_\phi, m_p,
m_n\right) &\sim&  {T}_{\gamma p \to \phi p} \left(m_\gamma, m_d - m
- m_n; m_\phi, m_p\right)
   C(1, m_d; l, m, 1, m_d - m) \nonumber \\
  &\times& C(1, m_d - m; 1/2, m_d - m - m_n, 1/2, m_n)      \psi_l(|{\mathbf{p}}_n|) Y_{l
m}(\Omega_{{\mathbf{p}}_n}), \label{Tp}
\end{eqnarray}
where  $m's$    stand for the spin projections of the particles and
$C's$ the CG coefficients, $Y_{l m}(\Omega_{\mathbf{p}})$  the
spherical harmonics. The wave function $\psi_l(|{\mathbf{p}}|)$,
describing the momentum-space radial distribution of the proton is
taken from Bonn potential.

For the case when the struck nucleon is a neutron, we first assume
that $T^{(n)}(P)=T^{(p)}(P)$ because Pomeron behaves like an
isoscalar. For the excitation of the postulated $N^*(3/2^-)$ from
neutron, we take the relativistic quark model of \cite{Capstick92}
as a guide. We assume that the photocoupling of the resonance is
similar to that of $N^*_{3/2^-}(2095)$ predicted in that model and
use its predicted helicity amplitudes to determine $g_{\gamma
NN^*}$. We then use  the fitted value of $g_{\gamma pN^*}g_{\phi
pN^*}$ determined in \cite{Kiswandhi10} and assume $g_{\phi nN^*}=
g_{\phi pN^*}$ to obatain $g_{\gamma nN^*}g_{\phi nN^*}$.

The IA amplitude with   FSI  included, i.e., where the struck
nucleon which produces the $\phi$ proceeds to interact with the
spectator nucleon, can be written as
\begin{eqnarray} &&<\vec q, \vec{p_1'},
\vec{p_2'}\mid T \mid k, \vec{p_1}, \vec{p_2} > \nonumber
\\
&=&\int d\vec{p_1''} <\vec q, \vec{p_1'}, \vec{p_2'}\mid
t_{NN}(E+i\epsilon) \mid \vec q, \vec{p_1''}, \vec{p_2}>\frac
{1}{E_i-H_0+i\epsilon}<\vec q, \vec{p_1''},
\vec{p_2}\mid t_{\gamma\phi} \mid k, \vec{p_1}, \vec{p_2} >\nonumber\\
&=&\int d\vec{p_1''} < \vec{p_1'}, \vec{p_2'}\mid
t_{NN}(E+i\epsilon) \mid   \vec{p_1''}, \vec{p_2}>\frac
{1}{E_i-H_0+i\epsilon}<\vec q, \vec{p_1''},
 \mid t_{\gamma\phi} \mid k, \vec{p_1}  >, \label{IA-FSI}
\end{eqnarray}
where $t_{NN}$ and the $t_{\gamma\phi}$ denote the two-body
amplitudes of $NN$ scattering and $\phi$ production from a single
nucleon, respectively. $H_0$ and $E_i$ refer to the free Hamiltonian
of the intermediate states and the initial energy of the $\gamma d$
system, respectively, and $E$ denotes the energy available to the
intermediate $NN$ pair before they rescatter.

Now note that the propagator of the intermediate states in Eq. (\ref{IA-FSI}) can be decomposed as,
\begin{equation}
\frac {1}{E_i-H_0+i\epsilon}=\frac {\bf P}{E_i-H_0} -i\pi\delta(E_i-H_0). \label{energy-shell}
\end{equation}
The amplitude $T$ in Eq. (\ref{IA-FSI}) obtained with the use of the
first and the second terms in Eq. (\ref{energy-shell}) as propagator
would give the FSI effects in which the $NN$ pair in the final state
scatter either off-energy-shell or on-shell.

The calculations with Eqs. (\ref{Tp}) and (\ref{IA-FSI}) are
tedious, time consuming, but manageable. We can now compare our
results with the data of \cite{Chang10,Chang10_DCS} and
\cite{Gao11}. However, a word of caution is in order concerning the
DCS presented in \cite{Chang10_DCS}. The DCS data shown in Fig. 2 of
\cite{Chang10_DCS} are actually the $d\sigma/dt$ of $\gamma
N\rightarrow \phi N$ extracted from $\gamma d\rightarrow \phi p n$,
namely, $(d\sigma_p/dt)\mid_{p} + \, (d\sigma_n/dt)\mid_{n}$, with
struck nucleon at rest, by the experimentalists. The extraction was
done with the help of GEANT3 which took into account experimental
parameters, Fermi motion, and the off-shell effects of the target
nucleons inside deuteron. With a prescription provided by Chang
\cite{Chang13}, we have reconstructed the raw data of
$d\sigma/dt(\gamma d\rightarrow \phi p n)$ for comparison with the
results of our calculations.

\begin{figure}[h]
\centering \sidecaption
\includegraphics[width=8.cm, height=5.4cm]{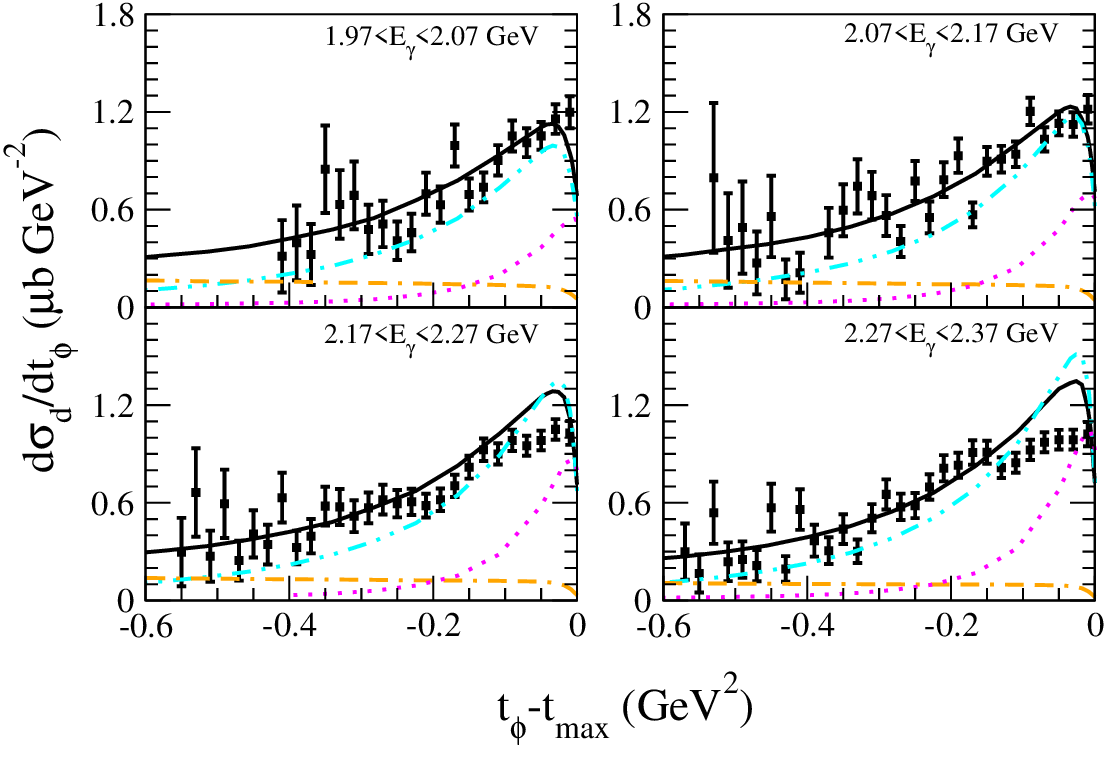}
\caption{Comparison of our model predictions for the DCS of $\gamma
d \to \phi p n$ as a function of momentum transfer $t_\phi$ at four
energy bins for four energy bins between  1.97 - 2.37 GeV  with data
of \cite{Chang10_DCS,Chang13}. See text for notations.}
\label{fig-5}       
\end{figure}
\vspace{-0.4cm}

Our model predictions for the DCS of $\gamma d \to \phi p n$ as a
function of momentum transfer $t_\phi=(q_\phi -k)^2$, where $q_\phi$
and $k$ are the momentum of $\phi$ and photon, respectively, at four
energy bins between 1.97 - 2.37 GeV are shown in Fig. \ref{fig-5}
and compared with the data of \cite{Chang10_DCS,Chang13}. The solid
lines are the full results of including nonresonant (NR), i.e,
Pomeron and $(\pi, \eta)$ exchanges, and resonance (R)
contributions, with $pn$ FSI taken into account. The dash-dotted,
and dash-dot-dotted lines are the results R and NR contributions
without the $pn$ FSI included, respectively, i.e., within the
impulse approximation. The dotted lines denote the results of FSI
effects of Eq. (\ref{IA-FSI}) only, namely, the contribution with
rescattering between $pn$ after $\phi$ is produced.  The squares
with error bars are the experimental data of Refs.
\cite{Chang10_DCS, Chang13}. One sees that 1). the resonance
contribution is nonnegligible;  2). FSI effect is large at forward
angles, i.e., $t_\phi-t_{max} \sim 0$. The agreement of the full
results are in general satisfactory but discrepancies are
considerable at forward angles.
\begin{figure}[ht]
\begin{minipage}{70mm}
\centering\includegraphics[width=5cm]{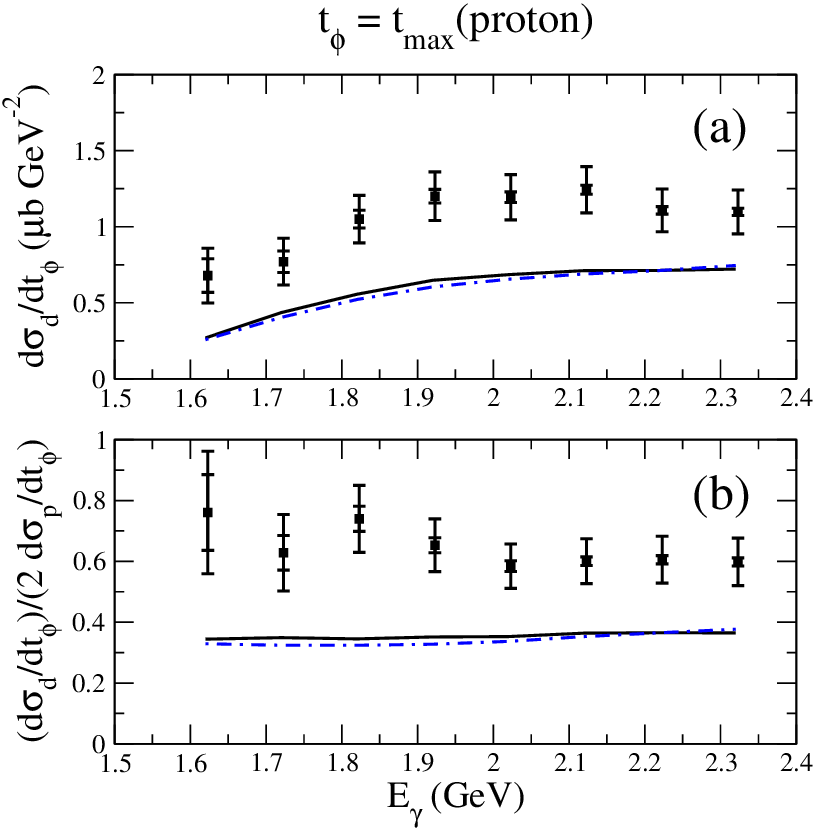} \caption{(a) The
DCS of $\gamma d \to \phi p n$ and (b) the ratio of the DCS of
$\gamma d \to \phi p n$ to twice the DCS of $\gamma p \to \phi p$,
both at forward angle as a function of $E_\gamma$.}\label{Fig8}
\end{minipage}
\hspace{\fill}
\begin{minipage}[ht]{65mm}
 {\hskip 0.5cm}In Fig. \ref{Fig8}, our predictions for (a) the DCS of $\gamma d \to
\phi p n$, and (b) the ratio of the DCS of $\gamma d \to \phi p n$
to twice the DCS of $\gamma p \to \phi p$, both at forward angle as
a function of $E_\gamma$, are compared with the LEPS data
\cite{Chang10_DCS,Chang13}. The solid and dash-dotted lines denote
the results of the full calculations and those without the inclusion
of FSI. It appears that the FSI effects are small. However, it
arises from an unexpected almost complete cancelation between the
on-shell and off-shell rescatterings effects as given in Eq.
(\ref{energy-shell}). The substantial discrepancies between our
 results and the data reflect the considerable difference between
 our prediction and the data already seen in Fig. \ref{fig-5}.
 In  our opinion, the recipe
\end{minipage}
\end{figure}

\noindent employed by LEPS to extrapolate their data
\cite{Chang10_DCS,Chang13}
 to the forward angles might be questionable as the DCS should show a
 sharp drop near forward direction \cite{Seki12} which is not seen in the LEPS data.

\begin{figure}[ht]
\begin{minipage}{75mm}
 {\hskip 0.5cm} Lastly, we compare our results with the CLAS DCS data
\cite{Gao11} for $1.65\le E_\gamma\le 1.75$ GeV as function of
$t_\phi-t_{max}$ in Fig. \ref{clas}. The notations are the same as
in Fig. \ref{fig-5}. Note that the CLAS data are taken in the much
larger momentum transfer  region as compared with the LEPS data. The
difference between our results and the data is substantial.

{\hskip 0.5cm} We have also calculated the spin density matrix
elements and compared them with the LEPS data \cite{Chang10}. The
agreement is in general reasonable and will be shown elsewhere
because of the space limit here.

{\hskip 0.5cm}In  summary, we have presented the results of a
calculation, within impulse approximation with final state
interaction between outgoing proton and neutron taken into account,
for the incoherent photoproduction of $\phi$ from deuteron at low
energies and compared them with the data
\end{minipage}
\hspace{\fill}
\begin{minipage}[h]{60mm}
\centering\includegraphics[width=4.5cm]{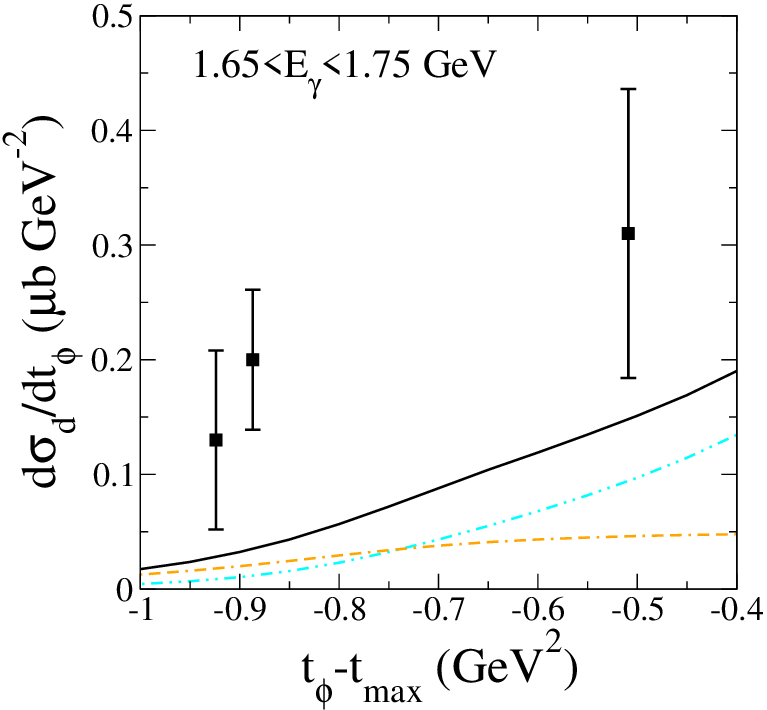}
\caption{Comparison of our results with the CLAS data for $1.65\le
E_\gamma\le 1.75$ GeV as function of $t_\phi-t_{max}$. Notation same
as Fig. \ref{fig-5}.}\label{clas}
\end{minipage}
\end{figure}
\vspace{-0.36cm} \noindent  from LEPS and CLAS.  The Fermi motion,
resonance, and the FSI effects are found  important in certain
kinematics. The overall agreement between our predictions and the
data ranges from poor to reasonable. Further improvements are
needed, e.g., the production of $\phi$ from the spectator nucleon by
intermediate mesons, like $\pi, \rho, \phi$, produced from the first
nucleon struck by the photon, and medium effects of the propagation
of the intermediate mesons, should be investigated. The last would
be to tune the strength of the photon and $\phi$ meson to excite a
neutron to produce the postulated resonance and see the possible
effect induced.

\section*{Acknowledgment}
This work is supported in part by the National Science Council of
ROC (Taiwan) under grant No. NSC101-2112-M002-025,  the National
Sciences Foundations of China (NSFC)  No.10975146 and 11035006. This
work is partly supported and by the DFG and NSFC through funds
provided to the Sino-German CRC 110.   Y.B.D.  thanks the Department
of Physics, National Taiwan university for the warm hospitality and
the financial support of Foundation of Chinese Development in
Taiwan.


\begin{thebibliography}{}
%
\bibitem{Kiswandhi10}A. Kiswandhi, J. J. Xie, and S. N. Yang, Phys. Lett. B {\bf 691}, 214 (2010);
A. Kiswandhi and S. N. Yang,  Phys. Rev. C {\bf 86}, 015203 (2012).
%
\bibitem{Chang10}W. C. Chang {\it et al.} (LEPS Collaboration), Phys. Rev. C {\bf 82}, 015205 (2010).
%
\bibitem{Chang10_DCS}W. C. Chang {\it et al.} (LEPS Collaboration), Phys. Lett. B {\bf 684}, 6-10 (2010).
%
\bibitem{Gao11} X. Qian {\it et al.} (CLAS Collaboration), Phys. Lett. B {\bf 696}, 338 (2011).

\bibitem{Capstick92} S. Capstick, Phys. Rev. D {\bf 46}, 2864 (1992).
%
\bibitem{Chang13}W. C. Chang, private communication.

%
\bibitem{Seki12} T. Sekihara, A. M. Torres, D. Jido, and E. Oset, Eur. Phys. J. A {\bf 48}: 10 (2012).
\end{thebibliography}
%
%

\end{document}